\documentclass[12pt,epsfig]{article}
\usepackage{epsfig}
\usepackage{amsmath}
\usepackage{amsthm}
\usepackage{enumerate}
\usepackage{mathrsfs}
\usepackage{dsfont}

\newtheorem{lemma}{Lemma}

\newtheorem{theorem}{Theorem}

\theoremstyle{definition}

\newtheorem{example}{Example}
\newtheorem*{example*}{Example}
\newtheorem*{remark*}{Remark}

\newfont{\boldlarge}{msbm10 scaled 1100}
\newfont{\boldsmall}{msbm10 scaled 800}
\newfont{\boldlarger}{msbm10 scaled 1200}
\newcommand{\RealF}{\mbox{\boldlarge R}}

\newcommand{\LNaturalF}{\mbox{\boldlarger N}}

\newcommand{\Expt}{\mbox{\boldlarge E}}

\newcommand{\Prob}{\mbox{\boldlarge P}}

\newcommand{\LProb}{\mbox{\boldlarger P}}

\newcommand{\eps}{\varepsilon}

\newcommand{\deff}{ \stackrel{\rm def}{=} }

\def\m{\mathcal}
\def\ind{\mathds{1}}

\begin{document}

\title{A Symbolic Dynamical System Approach to Lossy Source Coding with Feedforward}
\author{ Ofer Shayevitz \\
         {\sl \small University of California, San Diego} \\
         {\sl \small La Jolla, CA 92093, USA} \\
         {\sl \small ofersha@ucsd.edu}}

\date{ }
\maketitle


\begin{abstract}
It is known that modeling an information source via a symbolic dynamical system evolving over the unit interval, leads to a natural lossless compression scheme attaining the entropy rate of the source, under general conditions. We extend this notion to the lossy compression regime assuming a feedforward link is available, by modeling a source via a two-dimensional symbolic dynamical system where one component corresponds to the compressed signal, and the other essentially corresponds to the feedforward signal. For memoryless sources and an arbitrary bounded distortion measure, we show this approach leads to a family of simple deterministic compression schemes that attain the rate-distortion function of the source. The construction is dual to a recent optimal scheme for channel coding with feedback.
\end{abstract}

\thispagestyle{empty}

\section{Introduction}
Lossless compression of a discrete information source to its entropy rate $\m{H}$ is a well studied topic. A possibly lesser known approach to this problem is one based on symbolic dynamical systems, where the information generating mechanism is modeled by a randomly initialized iterative mapping of the unit interval to itself, and the emitted source sequence is a quantized observation of that process. For well behaved mappings the source sequence constitutes an \textit{expansion} of the initial point, i.e., corresponds to a unique such point. Furthermore, the prefixes of this expansion describe the initial point with (exponentially) increasing resolution, and the unit interval can be uniformly partitioned into $\approx 2^{n\m{H}}$ subintervals so that with high probability, the subinterval containing the initial point will have all its points admitting the same length-$n$ expansion. This leads to a conceptually simple and optimal compression scheme: A finite source sequence is mapped to a representing subinterval by computing the corresponding reverse trajectory of the dynamical system, and is reconstructed by following the trajectory of an arbitrary point in that subinterval\footnote{This has a flavor similar to arithmetic coding and (using variable-length coding) essentially coincides with it in some cases, see Example \ref{ex1}.}. A comprehensive study of the symbolic dynamics framework for information sources can be found in \cite{vallee01}. Some of the ideas can be traced back to R\'enyi, see \cite{halfant} and references therein.

In this paper, we extend the concept above to the lossy source coding regime, under the assumption that a noiseless feedforward link is available. This setting is described as follows: An encoder observes a stochastic source sequence $Y^n$ over some product alphabet $\m{Y}^n$, and maps it to a rate $R$ index set $E=\{1,2,\ldots, 2^{nR}\}$ using some encoding function $e:\m{Y}^n\mapsto E$. The index  is sent to the decoder. At time $k$, the decoder knows the sequence $Y^{k-1}$ via the feedforward link, and generates an approximation of $Y_k$ using a decoding function $f_k:E\times \m{Y}^{k-1}\mapsto \m{X}$, where $\m{X}$ is the reconstruction alphabet. The quality of the approximation is measured w.r.t. a distortion measure $d:\m{X}\times\m{Y}\mapsto\RealF^+$, by evaluating the time-averaged expected distortion:\vspace{-3pt}
\begin{equation*}
D = n^{-1}\sum_{k=1}^n\Expt\,d(f_k(e(Y^n),Y^{k-1}),Y_k)\vspace{-2pt}
\end{equation*}
The \textit{rate-distortion function} of the source is the infimum of all rates $R$ for which there exist encoding and decoding functions achieving a distortion at most $D$, for any $n$ large enough. It is denoted $R_{\textnormal{ff}}(D)$ under the feedforward assumption, and $R(D)$ where feedforward is absent (i.e., when restricting $f_k(e,y^{k-1}) = f_k(e)$).

This model has been initially motivated and studied in the context of competitive prediction \cite{weissman03}, where it was shown that feedforward does not decrease the rate-distortion function for a large family of sources (in particular, memoryless). An in-depth analysis of the rate-distortion function with feedforward appears in \cite{venkat07}. A simple scheme inspired by a successive error compression feedback coding technique and achieving the rate-distortion function for discrete memoryless sources, was suggested in \cite{matrinian04}. Another optimal protocol building on the Schalkwijk-Kailath scheme for channel coding with feedback over the AWGN, was suggested for the white Gaussian source \cite{Pradhan04}. In this paper, we suggest an alternative approach based in symbolic dynamics and motivated by a recent optimal feedback transmission scheme, termed \textit{posterior matching} \cite{posterior_matching}\cite{posterior_matching_isit08}\cite{posterior_matching_IT}. The suggested approach yields a conceptually simple compression protocol, which is shown to achieve the rate-distortion function for discrete memoryless sources with a bounded distortion measure.

\section{Preliminaries}
Random variables (r.v's) are denoted by upper-case letters, their realizations by corresponding lower-case letters. A r.v. $X$ (either real or discrete) is associated with a probability distribution $P_X(\cdot)$ (over $\RealF$, or over a discrete alphabet $\m{X}\subseteq\LNaturalF$) and we write $X\sim P_X$. The \textit{cumulative distribution function} (c.d.f.) of $X$ is denoted by $F_X$. We write $\Expt(\cdot)$ for expectation and $\Prob(\cdot)$ for the probability of an event within the parentheses. $H(X)$ is the entropy of a discrete r.v. $X$, $h(X)$ is the differential entropy of a continuous r.v. $X$, and $I(X;Y)$ is the mutual information between a pair of r.v. $X,Y$. We use $|\Delta|$ for the length of an interval $\Delta\subseteq\RealF$, $\log$ for $\log_2$, $\circ$ for function composition, $\overline{A}$ for the closure of the set $A$, $\ind_A(\cdot)$ for the indicator function over the set $A$,  $\m{I}$ for the open unit interval $(0,1)$, and $\m{I}_2 \deff \m{I}\times\m{I}$ for the open unit square. An \textit{open partition} of a set $A$  (in what follows, $\m{I}$ or $\m{I}_2$) is a family of disjoint open subsets $\{A_i\}$ of $A$, such that $\overline{\cup A_i} = \overline{A}$. A sequence $x^n$ over a finite alphabet is said to be (strongly) $\eps$-typical w.r.t. $P_X$, if the (zero order) empirical distribution of symbols in $x^n$ is $\eps$-close to the distribution $P_X$ in the supremum norm. The set of all such length $n$ sequences is denoted $\m{T}_{n,\eps}(P_X)$.

We now turn to define a (two-dimensional) \textit{dynamical source}, generalizing the definition in \cite{vallee01}. Note that in the sequel, we discuss in detail a significantly more restrictive family of dynamical sources. We provide the rather abstract definition below both for future reference, and as we believe it is more instructive.

A dynamical source $\m{S}$ has the following components:
\begin{itemize}
\item A triplet of alphabets $\m{X},\m{Y},\m{Z} \subseteq \LNaturalF$.
\item Two open partitions of $\m{I}$ into open intervals $\{\Pi^{0}_i\}_{i\in\m{X}}$ and $\{\Pi^{1}_j\}_{j\in\m{Z}}$, and the corresponding product partition $\Pi_{ij}=\Pi^{0}_i\times \Pi^{1}_j$ of $\m{I}_2$. Without loss of generality we assume that the intervals are arranged from left to right (or vice versa) according to the natural alphabet order.

\item Two functions $\sigma_{0}:\m{I}\mapsto\m{X}$, $\sigma_{1}:\m{I}\mapsto\m{Z}$ that are equal to $i,j$ over $\Pi^{0}_i,\Pi^{1}_j$ respectively.
\item A function $\xi:\m{X}\times\m{Z}\mapsto\m{Y}$, and its corresponding extension to $\zeta:\m{I}_2\mapsto\m{Y}$ that is constant and equal to $\xi(i,j)$ when restricted to $\Pi_{ij}$.
\item A mapping $T:\m{I}_2\mapsto\m{I}_2$ of the form
\begin{equation}\label{eq:T}
T(\theta,\phi) = (T_0(\theta,\zeta(\theta,\phi)), T_1(\theta,\phi))
\end{equation}
such that $T$ restricted to each $\Pi_{ij}$ is a continuous bijection, and $\left\{T(\Pi_{ij})\right\}_{(i,j)\in\xi^{-1}(k)}$ is an open partition of $\m{I}_2$ for each $k\in\m{Y}$.
\end{itemize}

Setting $(\theta_1,\phi_1)\in\m{I}_2$ as an \textit{initial state}, the source $\m{S}$ is associated with the following sequences, all of which are deterministic functions of the initial state:
\begin{itemize}
\item The \textit{state sequence} $(\theta^\infty,\phi^\infty)$ over $\m{I}_2$, recursively defined by $(\theta_n,\phi_n) = T(\theta_{n-1},\phi_{n-1})$.

\item The \textit{source sequence} $y^\infty$ over the alphabet $\m{Y}$, defined by $y_n = \zeta(\theta_n,\phi_n)$

\item The \textit{component sequences} $x^\infty,z^\infty$ over the alphabets $\m{X,Z}$ respectively, defined by \[x_n = \sigma_{0}(\theta_n)\,,\quad z_n = \sigma_{1}(\phi_n)\]

\end{itemize}
Furthermore, any finite source sequence $y^n$ corresponds to a \textit{fundamental set} $u_n(y^n)\subseteq\m{I}_2$, defined to be the set of all initial states $(\theta_1,\phi_1)\in\m{I}_2$ that result in the source sequence $y^n$.

Following \cite{vallee01} again, a \textit{probabilistic dynamical source} is a pair $(\m{S},P)$ where $\m{S}$ is a dynamical source, and $P$ is a probability measure equivalent to the Lebesgue measure over $\m{I}_2$ . Setting $(\Theta_1,\Phi_1)\sim P$ as the initial state, the source $(\m{S},P)$ is naturally associated with the \textit{stochastic} sequences $\Theta^\infty,\Phi^\infty,Y^\infty,X^\infty,Z^\infty$, all of which are deterministic functions of the initial state.

\section{Lossless Coding}
Let $(\m{S},P)$ be a probabilistic dynamical source with $|\m{Z}|=1$, i.e., one dimensional, and we can assume $\m{X}=\m{Y}$. In this case the fundamental sets are simply intervals in $\m{I}$ (in this section we disregard the redundant dimension). Under some further contraction conditions, an asymptotic equipartition property was shown to hold \cite{vallee01}, namely $n^{-1}\log|P(u_n(Y^n))|^{-1}$ tends in probability to the entropy rate $\m{H}(Y^\infty)$ of the source sequence. This immediately leads to an optimal compression protocol: The unit interval is uniformly partitioned into $\approx 2^{n(\m{H}+\delta)}$ representative intervals. The trajectory of the dynamical source is reversed using $Y^n$, namely recovering the fundamental interval  $T_0^{-1}(\cdot,Y_1)\circ\cdots\circ T_0^{-1}(\cdot,Y_{n-1})\circ T_0^{-1}(\m{I},Y_n)$. The index of a representative contained in the fundamental interval is used to describe the source sequence\footnote{If there is no such interval, an arbitrary index is used. This error event is of vanishing probability.}. To reconstruct $Y^n$, the dynamical source is initialized with any point inside the representative interval.

\begin{example}[Memoryless Sources]\label{ex1}
To generate a memoryless source over the alphabet $\m{Y}$, we set $P={\rm Uniform}(\m{I})$, $\xi(i,j) = \xi(i) = i$, and $T_0(\theta,i)$ to be affine and map $\Pi^0_i$ to $\m{I}$. This results in a source sequence that is i.i.d.-$P_Y$, where $P_Y(i) = |\Pi^0_i|$. If $T_0(\theta,i)$ are all monotonically increasing, the fundamental intervals are precisely those generated by the simple arithmetic coding protocol for the source, and coding them (the typical ones) as described (or alternatively, using a variable-rate code to obtain zero error) results in lossless compression with a rate approaching $H(Y)$. Note that in particular for $Y \sim {\rm Uniform} (\m{Y})$, the source sequence is simply the $|\m{Y}|$-base expansion of the initial state point.
\end{example}

\begin{example}[The Continued Fraction Source]
The continued fraction expansion of a number in $\m{I}$ can be generated by a dynamical source \cite{vallee01}. In this case we have $\m{Y} = \LNaturalF$, the open partition is $\Pi^0_i = (1\slash (i+1),1\slash i)$, $\xi(i,j) = \xi(i) = i$, and  $T_0(\theta,i) = \theta^{-1}-i = \theta^{-1}$ (mod $1$). Endowing the source with any probability measure $P$ that is equivalent to the Lebesgue measure over $\m{I}$, the state process converges to the invariant distribution that admits the density $f_{\rm inv}(\theta) = \frac{\log{e}}{1+\theta}\cdot\ind_{\m{I}}(\theta)$ \cite{contin_frac_book}.

Coding the fundamental intervals as described results in lossless compression with a rate approaching the entropy rate of the continued fraction source, which is given by $\m{H}(Y^\infty) = \frac{\pi^2\log{e}}{6}$. It is interesting to note that in this case, a more efficient (yet equivalent) coding mechanism for the fundamental intervals is readily available: Represent a finite source sequence $y^n$ by the unique rational number $\frac{p_n}{q_n}$ it is the continued fraction expansion of. It is well known that for almost all $\theta\in\m{I}$ (w.r.t. the Lebesgue measure), the denominator of the convergents of the continued fraction expansion satisfies $n^{-1}\log q_n \rightarrow \frac{\pi^2\log{e}}{12}$ \cite{contin_frac_book}, and so $(p_n,q_n)$ can be represented at a rate of twice this number, which is precisely the entropy rate of the continued fraction source.
\end{example}

\section{Lossy Coding with Feedforward}

\subsection{Motivation}
In the lossless setting, a finite source sequence was described by efficiently enumerating (typical) fundamental sets, obtained via a representation of an initial state up to a suitable resolution. In the lossy setting, we wish to provide only partial information regarding the fundamental set. To that end, a two-dimensional dynamical source model was introduced, where the high-level idea is to provide the decoder with a representation of the $\theta$-component of the initial state only. At time $k$, the decoder knows the sequence $Y^{k-1}$ (via feedforward), and can therefore compute the $\theta$-component $\Theta_{k-1}$ that corresponds to the initial state $\Theta_1$ it was given. This is made possible due to the restriction (\ref{eq:T}) on the structure of $T_0$, making its evolution dependent only on the $\theta$-component and the causal knowledge of the source sequence. Had it known the $\phi$-component as well, the decoder could have reconstructed $X_k,Z_k$ and hence $Y_k$. Here, it can only reconstruct $X_k$, which can serve as an estimate for $Y_k$.

So, our first task is, for a fixed source sequence distribution, to design a probabilistic dynamical source $(\m{S},P)$ that is consistent with this distribution, and also makes $X^n,Y^n$ dependent in a prescribed way so that this reconstruction has low distortion. However, there is an even more difficult obstacle. The initial $\theta$-component has to be described with a finite rate, and (loosely speaking) this should be done while making sure that an initial $\phi$-component can be selected so that the statistical dependence above is roughly maintained. For memoryless sources, both tasks can be accomplished.

\subsection{Memoryless Sources}
Let $P_Y$ be a probability distribution over the alphabet $\m{Y}$. There are many different probabilistic dynamical sources for which the source sequence is i.i.d.-$P_Y$. One simple example was given in the previous section, where $|\m{Z}|=1$ and $T$ is affine on any $\Pi^0_i$, and corresponds to a lossless compression with rate $H(Y)$. However, in two dimensions there is an abundance of distinct probabilistic dynamical sources that admit an i.i.d.-$P_Y$ source sequence.

Consider any channel $P_{X|Y}$ from  $Y$ to $X$ over the alphabets $\m{X}\times\m{Y}$, let $P_{XY} = P_Y\times P_{X|Y}$ be the joint distribution and let $P_{Y|X}$ be the corresponding test channel from $X$ to $Y$. The following Lemma is easily observed \cite{El-Gamal--Kim2009}.
\begin{lemma}\label{lem:func}
There exists an alphabet $\m{Z}$ of size $|\m{Z}|\leq |\m{X}|(|\m{Y}|-1)+1$, a function $\xi:\m{X}\times\m{Z}\mapsto \m{Y}$, and a r.v. $Z$ independent of $X$, such that $(X,\xi(X,Z))\sim P_{XY}$.
\end{lemma}

Now, let us define the following dynamical source $\m{S}$. The construction is motivated by the \textit{posterior matching scheme}, a capacity achieving feedback transmission scheme for memoryless channels with feedback \cite{posterior_matching}\cite{posterior_matching_isit08}\cite{posterior_matching_IT}.
\begin{itemize}
\item $|\Pi^0_i| = P_X(i)$ for any $i\in\m{X}$.
\item $|\Pi^1_j| = P_Z(j)$ for any $j\in\m{Z}$.
\item The function $\xi$ is that of Lemma \ref{lem:func}, $\zeta$ its natural extension.
\item The mapping $T=(T_0,T_1)$ is defined as follows:
    \begin{itemize}
    \item Let $F_{X|Y}$ be the conditional c.d.f. for $P_{X|Y}$. For any fixed $k\in\m{Y}$, $T_0(\theta,k)$ is a continuous non-decreasing function from $\m{I}$ onto $\m{I}$, is affine on each $\Pi^0_i$, and is equal to $F_{X|Y}(i|k)$ on the right edge of $\overline{\Pi^0_i}$.
        \item $T_1(\theta,\phi) = T_1(\phi)$ is one dimensional, affine on each $\Pi^1_j$ and maps it onto $\m{I}$.
    \end{itemize}
\end{itemize}

\begin{remark*}
Note that when $P_{X|Y}$ is noiseless (e.g., $X=Y$) then $\m{S}$ collapses to the one dimensional lossless construction of Example \ref{ex1}.
\end{remark*}

\begin{lemma}\label{lem:fund_set}
For any $y\in\m{Y}^n$, the fundamental set $u_n(y^n)$ of the dynamical source $\m{S}$ is a finite disjoint union of product rectangles. The projections of these rectangles onto the $\theta$-axis form a set of at most $n(|\m{X}|-1)+1$ distinct intervals.
\end{lemma}
\begin{proof}
The first assertion follows easily from the affinity of $T$. For $n=1$, the number of distinct intervals on the $\theta$-axis is exactly $|\m{X}|$. For any fixed $k\in\m{Y}$, $T_0(a,k)$ is quasi-affine over $\m{I}$ as a function of $a$, with at most $|\m{X}|-1$ corner points. Hence the number of distinct intervals can increase by at most $|\m{X}|-1$ at each step.
\end{proof}

The following Lemma is adapted from \cite{posterior_matching_IT}.
\begin{lemma}\label{lem:pm}
Let $P\sim{\rm Uniform}(\m{I}_2)$. The probabilistic dynamical source $(\m{S},P)$ has the following properties:
\begin{enumerate}[(a)]
\item The sequence $Z^\infty$ is i.i.d.-$P_Z$, $Z_n$ is statistically independent of $X^n$.
\item $(X_n,Y_n)\sim P_{XY}$, $X_n$ is statistically independent of $Y^{n-1}$
\item The source sequence $Y^\infty$ is i.i.d.-$P_Y$, and  $\,Y_n-X_n-X^{n-1}Y^{n-1}Z^{n-1}$ form a Markov chain.
\item $I(\Theta_1;Y^n) = nI(X;Y)$
\end{enumerate}
\end{lemma}
\begin{proof}
Assertion (a) is immediate: $Z^\infty $ is a deterministic function of $\Phi_1$ and evolves according to the memoryless dynamical law described in Example \ref{ex1}, hence is i.i.d-$P_Z$. Furthermore, $X^n$ is a deterministic function of $(\Theta_1,Z^{n-1})$ which are mutually independent of $Z_n$. For the other assertions, see \cite{posterior_matching_IT}.
\end{proof}

Define
\begin{equation}\label{eq:dec_set}
\omega_k(\cdot) \deff T_0^{-1}(\cdot,k)\,,\quad \Delta_n^\eps(y^n) \deff \omega_{y_{1}}\circ\cdots\circ\omega_{y_{n-1}}\circ\omega_{y_n}((\eps,1-\eps))
\end{equation}
Namely, the interval $\Delta^\eps_n(y^n)$ is obtained by reversing the trajectory of the (edges of the) interval $(\eps,1-\eps)$. The following result, also adapted from \cite{posterior_matching_IT}, is central to our derivations.

\begin{theorem}\label{thrm:pm}
Suppose that $P_{XY}$ is strictly positive over $\m{X}\times \m{Y}$, and that for any fixed $\theta\in\m{I}$, $T_0(\theta,j)$ is not a constant function of $j$. Then for any $\eps>0$,
\begin{enumerate}[(a)]
\item ${\displaystyle\lim_{n\rightarrow\infty}}\LProb(|\Delta^\eps_n(Y^n)| > 2^{-nR}) = 0$ for any $R < I(X;Y)$.
\item ${\displaystyle\lim_{n\rightarrow\infty}}\LProb(\Theta_1\not\in\Delta^\eps_n(Y^n)) \leq 2\eps$.
\end{enumerate}
\end{theorem}
Loosely speaking, Theorem \ref{thrm:pm} implies that by observing the source sequence, the initial $\Theta_1$ component of the state sequence can be found up to a resolution of $2^{-nI(X;Y)}$. In a feedback communication setting, this initial value represents a \textit{message} to be sent over the channel $P_{Y|X}$, and this concentration result means that one can reliably transmit roughly $2^{nI(X;Y)}$ such messages and decode them with high reliability, which corresponds to a communication rate of \textit{at most} $I(X;Y)$ bits per channel use. In order to be able to generate $X_n$ (channel input) the encoder needs to know the $Y^{n-1}$ on top of the message $\Theta_1$, hence the feedback. In the dual lossy source coding with feedforward setting we consider, $\Theta_1$ plays the role of a lossy description of the source sequence, and we will need \textit{at least} $I(X;Y)$ bits per source symbol to represent it with high enough accuracy. In order to be able to generate $X_n$ (lossy reconstruction of $Y_n$) the decoder needs to know the $Y^{n-1}$ on top of the (quantized representation of the) lossy description $\Theta_1$, hence the feedforward.

Fix the block size $n$, and set $R = I(X;Y)+\delta$ for some $\delta>0$. Let $\{\m{J}_m\}_{m=1}^{\lfloor 2^{nR}\rfloor}$ be an open partition of $\m{I}$ into equi-sized intervals, and let $a_m$ be the midpoint of $\m{J}_m$. Denote the set of all midpoints by $A_n$.
\begin{lemma}\label{lem:typical}${\displaystyle\lim_{n\rightarrow\infty}\LProb \Big{(}\bigcup_{\stackrel{\scriptstyle\theta\in A_n\cap \Delta_n^\eps(Y^n)}{(\theta,\phi)\in u_n(Y^n)}}\hspace{-15pt}\left\{(x^n(\theta,\phi),z^n(\theta,\phi))\right\}\,\cap\, \m{T}_{n,\delta}(P_{XZ}) = \emptyset\Big{)} = 0}$.
\end{lemma}
\begin{proof}[Proof Outline]
For lack of space we only describe the main elements of the proof, skipping some details. Let $V(y^n)$ be the set of indices $m$ such that $a_m\in\Delta_n^\eps(y^n)$, and $\m{J}_m$ intersects with two or more intervals that are projections of a product rectangle in $u_n(y^n)$ onto the $\theta$-axis. By Lemma \ref{lem:fund_set}, $|V(y^n)|\leq n|\m{X}|$. Define
\[q_n(y^n) \deff P_{\Theta_1|Y^n}\Big{(}\bigcup_{m\in V(y^n)} \m{J}_m\,\mid\, y^n\Big{)}\]
and consider Theorem \ref{thrm:pm} with a rate $I(X;Y)-\eps_1$. Now, assume to the contrary that $\LProb\left(q_n(Y^n)>\eps_2\right) > \eps_3$ for some fixed $\eps_2\in(2\eps,1-2\eps), \eps_3>0$, i.e., with probability at least $\eps_3$ the distribution of $\Theta_1$ given $Y^n$ has a mass at least $\eps_2$ inside that polynomial sized set of intervals. Then we have (some transitions assuming $n$ large enough)
\begin{align}\label{eq:conc}
\nonumber n^{-1}I(\Theta_1;Y^n) &= -n^{-1}h(\Theta_1|Y^n)
\\
\nonumber &\geq n^{-1}\eps_3\left[\eps_2\log\left({\eps_2\cdot 2^{nR}\cdot(n|\m{X}|)^{-1}}\right)\right.
\\
\nonumber &\hspace{0.5cm}+ \left.(1-2\eps-\eps_2)\log\left((1-2\eps-\eps_2)\cdot2^{n(I(X;Y)-\eps_1)}\right)\right]
\\
\nonumber  &\hspace{0.5cm}+n^{-1}(1-\eps_3)(1-2\eps)\log((1-2\eps)\cdot 2^{n(I(X;Y)-\eps_1)})
\\
& = (1-2\eps)\cdot I(X;Y) + \delta\eps_2\eps_3 - \eps_1(1-2\eps-\eps_2\eps_3) + O(\log{n}\slash n)
\end{align}
where we have used the concentration result of Theorem \ref{thrm:pm} for the inequality transition. Since $\eps,\eps_1$ can be taken arbitrarily small for $n$ large enough, the right-hand-side of (\ref{eq:conc}) can be made larger than $I(X;Y)$, contradicting Lemma \ref{lem:pm}. Note that this argument is similar in essence to the converse to the channel coding Theorem \cite{cover}.

We conclude that $q_n(Y^n)\rightarrow 0$ in probability, which loosely speaking means that with high probability, $P_{\Theta_1|Y^n}$ is mostly concentrated on $\Delta_n^\eps(y^n)\setminus \cup_{m\in V(y^n)} \m{J}_m$ for large $n$. Using typicality arguments together with the properties in Lemma \ref{lem:pm}, this can be shown to imply that with high probability we can find $\theta$ in that set together with some $\phi$ such that $(\theta,\phi)\in u_n(Y^n)$ and $(x^n(\theta,\phi),z^n(\theta,\phi))\in\m{T}_{n,\delta}(P_{XZ})$. By definition, $\theta\in\m{J}_m$ where $\m{J}_m$ is a subset of some interval which is a projection of a product rectangle in $u_n(Y^n)$. This is turn implies that $x^n(a_m,\phi) = x^n(\theta,\phi)$ and $z^n(a_m,\phi) = z^n(\theta,\phi)$, concluding the proof.
\end{proof}

We are now ready to describe the compression protocol.
\vspace{0.1cm}

\hspace{-0.6cm}\underline{Encoder}
\begin{enumerate}[(a)]
\item Given the sequence $y^n$, compute $\Delta_n^\eps(y^n)$ using the recursion (\ref{eq:dec_set}).
\item \label{item:enc}Out of the $\approx 2^{n\delta}$ intervals $\m{J}_m\subseteq\Delta_n^\eps(y^n)$, find the one with the least index\footnote{It seems that a random selection should work with high probability, making the process simpler. However, this was not verified.} $m$, for which there exists $\phi\in\m{I}$ such that $(x^n(a_m,\phi),z^n(a_m,\phi))\in\m{T}_{n,\eps}(P_{XZ})$. If no such index exists, arbitrarily set $m=1$.
\item Send the index $m$ to the decoder, which requires a rate of $I(X;Y)+\delta$ bits per source symbol.
\end{enumerate}

\hspace{-0.6cm}\underline{Decoder}
\begin{enumerate}[(a)]
\item Initialization: Set $\theta_1=a_m$, compute $x_1=\sigma_0(\theta_1)$.
\item For any $k$, predict $\widehat{y}_k = x_k$\label{repstep1}.
\item Receive the true $y_k$ via the feedforward link, compute $\theta_{k+1} = T_0(\theta_k,y_k)$ and $x_{k+1} = \sigma_0(\theta_{k+1})$\label{repstep2}.
\item Repeat steps (\ref{repstep1})--(\ref{repstep2}) up to $k=n$.
\end{enumerate}

The compression rate attained by the scheme is $R=I(X;Y)+\delta$. If encoding step (\ref{item:enc}) is successful then the pair $(x^n,z^n)$ is jointly $P_{XZ}$-typical, which implies that $(x^n,y^n)$ is jointly $P_{XY}$-typical. By Lemma \ref{lem:typical}, when encoding an i.i.d-$P_Y$ sequence $Y^n$ this occurs with probability approaching $1$ as $n\rightarrow \infty$. Since the distortion measure is bounded, the expected distortion achieved by the scheme is given by $D=\Expt_{P_{XY}} d(X,Y) + o(1)$.

The development above holds for any $P_Y$ and $P_{X|Y}$ that satisfy the requirements of Theorem \ref{thrm:pm}. The strict positivity constraint for $P_{XY}$ has a negligible effect, since such distributions can always be approximated arbitrarily via admissible distributions, and the distortion measure is bounded. The second constraint is redundant as it can always be averted by using a variant of the the probabilistic dynamical source, as in the channel coding case \cite{posterior_matching_IT}\cite{posterior_matching_allerton}. Hence, we have proved the following result.

\begin{theorem}
For any discrete memoryless source and bounded distortion measure, the protocol described above can perform arbitrarily close to the rate distortion function of the source.
\end{theorem}

\begin{figure}
  \begin{center}
    \epsfig{file=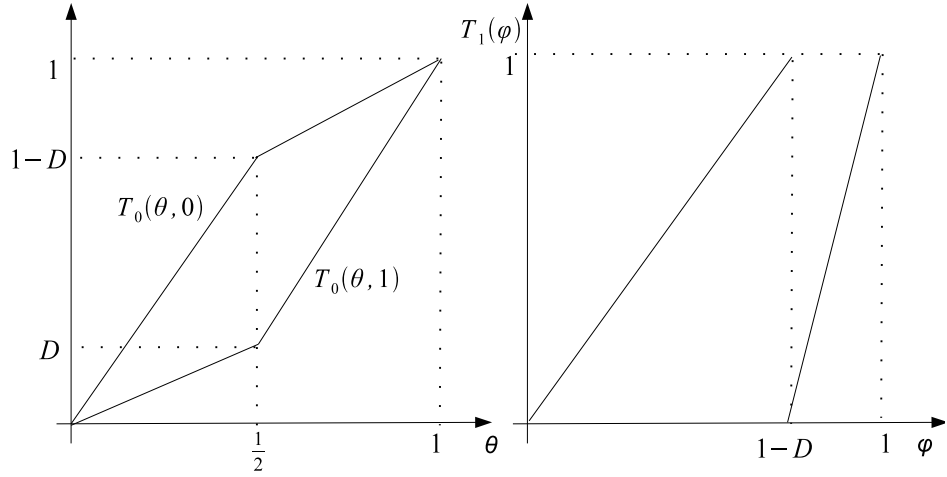, scale = 0.5}
    \caption{The mapping $T=(T_0,T_1)$ for Bern($\frac{1}{2}$) source, Hamming distortion $D$}
    \label{fig:T}
  \end{center}
\end{figure}

\begin{figure}
  \begin{center}
    \epsfig{file=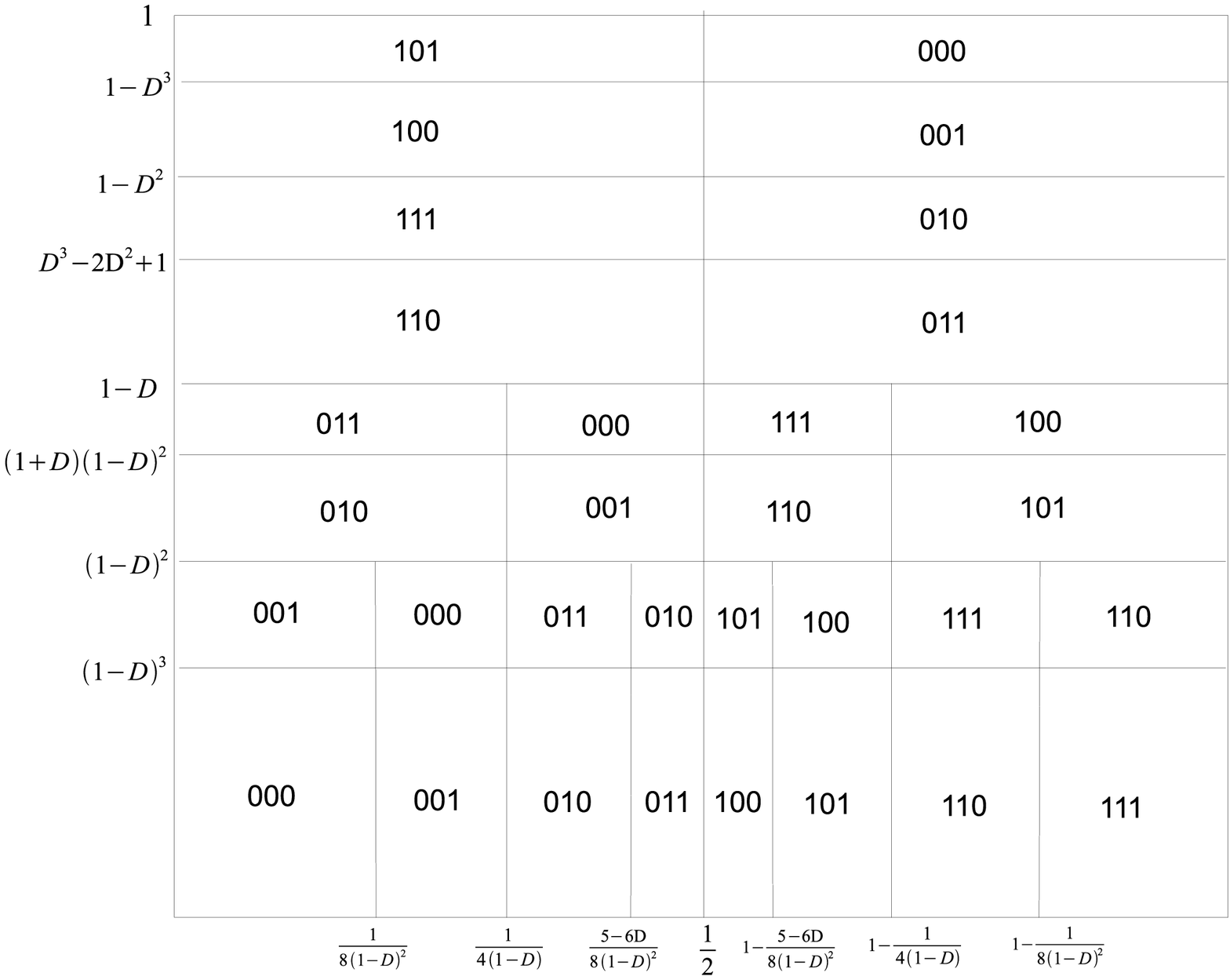, scale = 0.5}
    \caption{Fundamental sets with $n=3$ for Bern($\frac{1}{2}$) source, Hamming distortion $D$}
    \label{fig:fund_sets}
  \end{center}
\end{figure}

\begin{example}[Bernoulli Source and Hamming Distortion]
Let $\m{X}=\m{Y} = \{0,1\}$, $Y\sim{\rm Bern}(\frac{1}{2})$, $d(\cdot,\cdot)$ the Hamming distortion measure. The rate distortion function $R_{\textnormal{ff}}(D)=R(D)=1-h_b(D)$ is achieved by $X\sim{\rm Bern}(\frac{1}{2})$, $Z\sim{\rm Bern}(D)$ independent of $X$, and $Y=X+Z$ (mod $2$). The partitions and mappings are given by
\[\Pi^0_0 = (0,\frac{1}{2})\,\quad\Pi^0_1 = (\frac{1}{2},1)\]
\[\Pi^1_0 = (0,1-D)\,\quad \Pi^1_1 = (1-D,1)\]
\[T_0(\theta,0) = 2\theta (1-D)\cdot\ind_{\Pi^0_0}(\theta) + (2D\theta + 1-2D)\cdot\ind_{\Pi^0_0}(\theta)\]
\[T_0(\theta,1) = 2\theta D\cdot\ind_{\Pi^0_0}(\theta) + (2(1-D)\theta + 2D-1)\cdot\ind_{\Pi^0_0}(\theta)\]
\[T_1(\phi,k) = T_1(\phi) = \frac{\phi}{1-D}\cdot\ind_{\Pi^1_0}(\phi) + \frac{\phi-(1-D)}{D}\cdot\ind_{\Pi^1_1}(\phi) \]
The mappings and the fundamental sets for $n=3$  are depicted in Figures \ref{fig:T} and \ref{fig:fund_sets}.
\end{example}

\section{Conclusions}
A symbolic dynamical system approach to lossy source coding with feedforward was introduced, yielding in particular a conceptually simple and optimal compression protocol for memoryless sources. In this latter case, the construction is dual to the posterior matching feedback communication scheme for memoryless channels. Future work should examine the suggested framework for sources with memory. A reasonable first goal could be the case where the $\phi$-component of the dynamical source evolves independently as in the memoryless case, yet generates e.g. a Markovian $Z^n$.

\bibliographystyle{IEEEbib}

\begin{thebibliography}{10}

\bibitem{vallee01}
B.~Vall\'ee,
\newblock ``Dynamical sources in information theory: Fundamental intervals and
  word prefixes,''
\newblock {\em Algorithmica}, vol. 29, pp. 262--306, February 2001.

\bibitem{halfant}
M.~Halfant,
\newblock ``Analytic properties of {R}\'enyi's invariant density,''
\newblock {\em Israel Journal of Mathematics}, vol. 27, no. 1, 1977.

\bibitem{weissman03}
T.~Weissman and N.~Merhav,
\newblock ``On competitive prediction and its relation to rate-distortion
  theory,''
\newblock {\em IEEE Trans. Info. Theory}, vol. 49, no. 12, pp. 3185--3194, Dec.
  2003.

\bibitem{venkat07}
R.~Venkataramanan and S.~Sandeep Pradhan,
\newblock ``Source coding with feed-forward: Rate-distortion theorems and error
  exponents for a general source,''
\newblock {\em Information Theory, IEEE Transactions on}, vol. 53, no. 6, pp.
  2154--2179, June 2007.

\bibitem{matrinian04}
E.~Martinian and G.W. Wornell,
\newblock ``Source coding with fixed lag side information,''
\newblock in {\em Proceedings of the 42th Allerton Conference on Communication,
  Control, and Computing}.

\bibitem{Pradhan04}
S.~Sandeep Pradhan,
\newblock ``Source coding with feedforward: Gaussian sources,''
\newblock in {\em Information Theory, 2004. ISIT 2004. Proceedings.
  International Symposium on}, June-2 July 2004.

\bibitem{posterior_matching}
O.~Shayevitz and M.~Feder,
\newblock ``Communication with feedback via posterior matching,''
\newblock in {\em Proc. of the International Symposium on Information Theory},
  2007.

\bibitem{posterior_matching_isit08}
O.~Shayevitz and M.~Feder,
\newblock ``The posterior matching feedback scheme: Capacity achieving and
  error analysis,''
\newblock in {\em Proc. of the International Symposium on Information Theory},
  2008.

\bibitem{posterior_matching_IT}
O.~Shayevitz and M.~Feder,
\newblock ``Optimal feedback communication via posterior matching,''
\newblock submitted to IEEE Trans. Info. Theory, available online at
  arXiv:0909.4828 [cs.IT].

\bibitem{contin_frac_book}
M.~Iosifescu and C.~Kraaikamp,
\newblock {\em Metrical Theory of Continued Fractions},
\newblock Kluwer Academic Publishers, 2002.

\bibitem{El-Gamal--Kim2009}
A.~El Gamal and Y-.H Kim,
\newblock {\em Lecture Notes on Network Information Theory},
\newblock Stanford University and UCSD, 2009.

\bibitem{cover}
T.M. Cover and J.A Thomas,
\newblock {\em Elements of Information Theory},
\newblock John Wiley \& Sons, Inc., 1991.

\bibitem{posterior_matching_allerton}
O.~Shayevitz,
\newblock ``Posterior matching variants and fixed-point elimination,''
\newblock in {\em Proceedings of the 47th Allerton Conference on Communication,
  Control, and Computing}.

\end{thebibliography}

\end{document}